\documentclass[twocolumn,showpacs,preprintnumbers,amsmath,amssymb]{revtex4}

\usepackage{multirow}%
\usepackage{graphicx}
\usepackage{dcolumn}
\usepackage{bm}

\newcommand{\ybin}{YbInCu$_{4}$}
\newcommand{\urusi}{URu$_{2}$Si$_{2}$}
\newcommand{\beg}{\begin{equation}}
\newcommand{\en}{\end{equation}}

\begin{document}

\title{Kondo lattice excitation observed using resonant inelastic X-ray scattering at the Yb $M_5$ edge}

\author{J. N. Hancock$^{1}$}
\author{M. Dzero$^{2}$}
\author{J. Sarrao$^{3}$}
\author{T. Schmitt$^{4}$}
\author{V. Strocov$^{4}$}
\author{M. Guarise$^{5}$}
\author{M. Grioni$^{5}$}

\affiliation{$^{1}$ Department of Physics and Institute for Materials Science, University of Connecticut, Storrs, Connecticut, 06269 USA}
\affiliation{$^{2}$Department of Physics, Kent State University, Kent, OH, 44242, USA}
\affiliation{$^{3}$Los Alamos National Laboratory, Los Alamos, New Mexico 87545, USA}
\affiliation{$^{4}$Paul Scherrer Institut, CH-5232, Villigen, PSI, Switzerland}
\affiliation{$^{5}$\'Ecole Polytechnique F\'ed\'ederale de Lausanne, Institut de Physique des Nanostructures, CH 1015 Lausanne, Switzerland}

\date{\today}

\begin{abstract}
We present a study of the resonant inelastic scattering response of \ybin\ excited at the tender Yb $M_5$ X-ray edge. In the high-temperature, paramagnetic phase, we observe a multiplet structure which can be understood at an ionic level. Upon cooling through the valence transition at $T_v\sim$ 40$K$, we observe a strong renormalization of the low-energy spectra, indicating a sensitivity to the formation of an intermediate valence phase at low temperatures. Similar spectrum renormalization has been observed in the optical conductivity, which suggests that the low-energy electronic structure 
possesses both mixed conduction and localized character.
\end{abstract}

\pacs{PACS}

\maketitle
\section{Introduction}

Materials which contain elements with intermediate valence configurations have remained in the focus of intense research for several decades now. They host a rich variety of physical phenomena including quantum critical phase transitions, unconventional superconductivity and topologically protected metallic behavior localized on sample boundaries. Years of research have provided ample evidence that these phenomena are driven by strong interactions between the predominately localized electronic degrees of freedom and their admixture with itinerant states. As such, the valence states available to the localized species are crucial for low-energy behavior of these correlated systems. Instabilities related to changes in the electronic valence configurations occur across the $f$-filling series of elements, and particularly simple limits are realized in Ce- and Yb-based materials. In cerium materials, valence fluctuations are realized between $f^0$ and $f^1$ states, differing by a single $f$ electron, while in Yb, fluctuations between $f^{13}$ and $f^{14}$ valence states differ by one $f$ hole. 

For elemental cerium metal, the pressure-induced $\alpha$-$\gamma$ transition appears to be described by a collapse of valence states fueled by the strong dependence of hybridization and volume on valence state\cite{Allen1982,Lanata2013}. Although involving a strong change in lattice volume, the transition appears isostructural, compelling an explanation in purely electronic terms. This fascinating situation has been the topic of recent research in correlated systems, but is not an isolated case. YbInCu$_4$, for example, is a $f-$hole analog of this famous problem, featuring another 1st order valence transition driven by an experimentally accessible temperature-change at $T_v$$\simeq$40K. This transition is also iso-structural, with a large change in screening strength quantified by an effective Kondo temperature $T_K$$\simeq$ 17K at $T>T_v$ and an order of magnitude higher $T_K$$\simeq$400 K at $T<T_v$. 
Intriguingly, the valence transition can be fully suppressed by an external magnetic field and in this
sense is analogous to the iso-structural valence transition in Ce. Furthermore, the line separating phases with two different valence states remains universal: $(T/T_{v})^2+(B/B_v)^2=1$ (here $B_v$ is the value of magnetic field corresponding to a phase transition at zero temperature while $T_v$ is a critical temperature at zero field)\cite{Dzero2000}, suggestive of universal and tractable low-energy physics stemming from a single energy scale. Valence transitions of this type, wherein local-moment and screened-moment magnetism compete, are very appealing cases of more general class of problems in correlated electron systems.

\begin{figure}
\begin{center}
\includegraphics[width=3.2in]{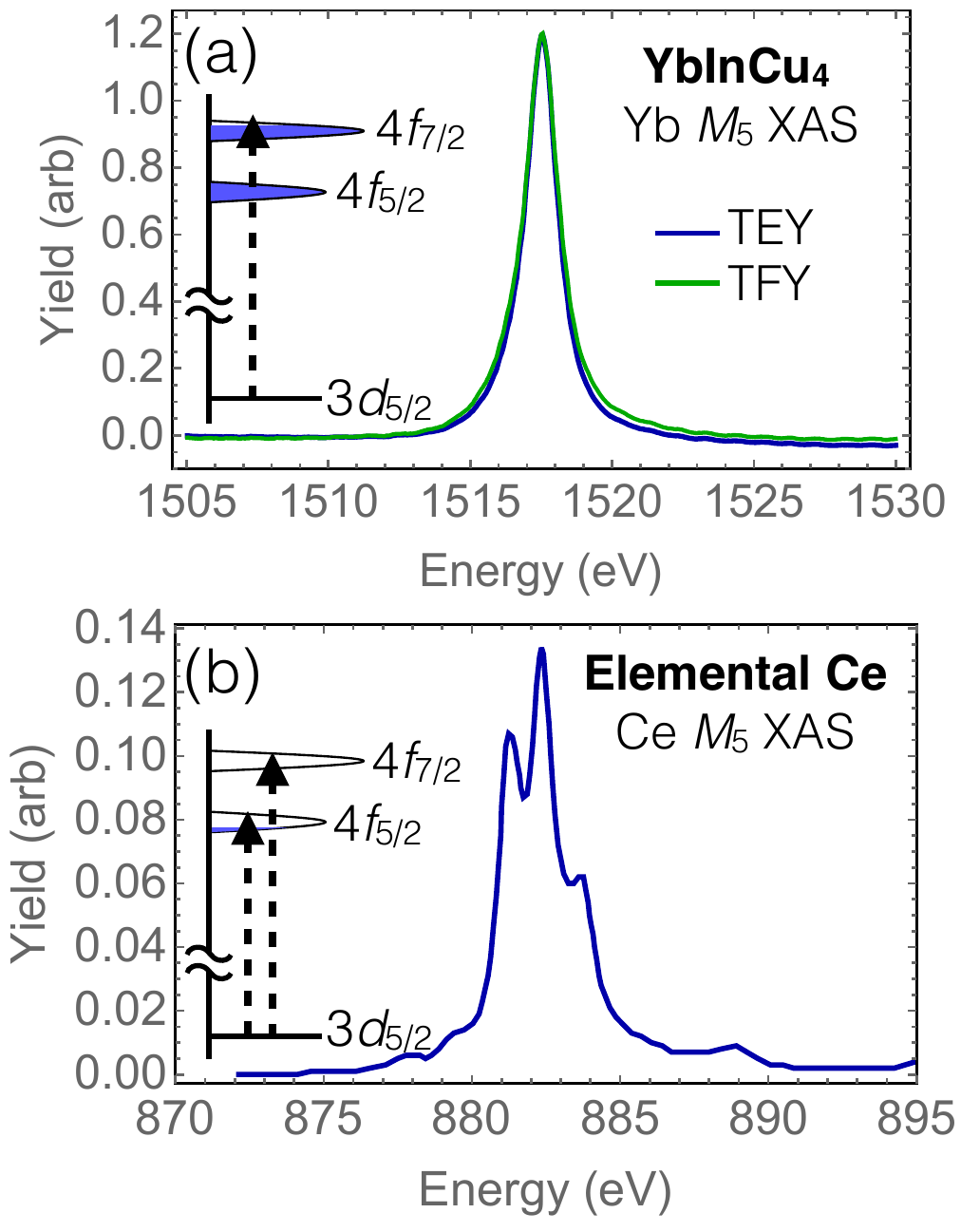}
\caption{(a) Total electron yield (TEY) and total fluorescence yield (TFY) of YbInCu$_4$ at the Yb $M_5$ resonance. The inset shows schematically the density-of-states profile and corresponding electronic transition from the 3$d_{\frac{5}{2}}$ core electronic level to the unoccupied portions of the 4$f$ manifold, which results in a single resonance for the Yb filling factor. (b) shows in contrast that Ce-based compounds have a far more complex resonance profile due to the largely unoccupied 4$f$ manifold. From \cite{Howald2015}.
}
\label{ }
\end{center}
\end{figure}


Upon substitution of transition metals YbXCu$_4$ (X=Ag, Mg, Zn) the valence transition does not appear. Furthermore, the low-$T$ properties of these compounds display behavior typical of members of the heavy fermion class of systems and can be thought of as being adiabatically connected to the low-$T$ phase of the \ybin\ system \cite{Hancock2004b,hancock06a,corn98}. Theoretical modeling of these compounds reveal the presence of a pseudo-gap in the itinerant electronic band structure. In \ybin,\ however, this significant reduction in the density of states occurs exceptionally near the Fermi level - a feature which was experimentally observed recently using hard x-ray resonant inelastic X-ray scattering (RIXS) at the Yb $L_3$ edge\cite{Jarrige2015} along with concomitant changes in this pseudo-gap and associated RIXS-induced electronic transitions. It has been long suggested \cite{Figueroa1998} that $\rho_F$ in this system is highly textured near the Fermi level, giving another influence that can control the valence state in YbInCu$_4$. In this scenario, the high temperature, integer/trivalent, paramagnetic state has a low $\rho_F$ and insufficient free carriers to screen the $J$=7/2 Yb moment. The transition is to a mixed valent state with 0.8 $f$ holes, requiring a lowering of the Fermi level into the region of larger DOS which can effectively screen the moment and release the corresponding magnetic entropy $\sim R\ln 8$\cite{Sarrao1998}. YbInCu$_4$ then becomes an interesting case where the Kondo scale and degree of screening is switched by temperature, in contrast to single impurity results and archetypal heavy fermion materials.

Here we use the peculiar Kondo-switching behavior of YbInCu$_4$ to explore broadly the response of Kondo lattice systems to the RIXS probe through its application at the tender Yb-$M_5$  x-ray edge. The benefits of studying the valence transition in the Yb compounds are multi-fold. Figure 1a shows the experimentally-obtained X-ray absorption spectrum (XAS) through the energy of the Yb $M_5$ edge. We observe a single Lorentzian shape peak, free from sidebands and multi-valence and spin-orbit resonances (Fig 1a)\cite{Amorese2016}. In contrast, elemental Ce shows many peaks in absorption, obfuscating the resonance structure below the core level broadening (Fig 1b). The Yb $M$ edges are accessible in the new suite of soft X-ray beamlines with improved energy resolution, but remain completely unexplored to our knowledge. In this pilot study of the RIXS response across a valence transition in YbInCu$_4$, we observe low-energy excitations identified previously in infrared studies and assigned to a hybridization gap\cite{Garner2000,DeGiorgi2001,Dordevic2001,Hancock2004b,okamura2007}. Our observation constitutes a demonstration of how the RIXS technique can be used to study the momentum-space texture of spin and charge excitations on the energy scales involved in the composite quasiparticles of heavy fermion and related systems.

\section{Experimental details}
Single crystals were grown using the flux method, and gave a transition temperature width $\Delta T_v$ $< $1$K$ soon after growth\cite{Hancock2004a,Sarrao1999}. Magnetometry reveals that over time, the transition temperature reduced by 1K, with slight rounding of the susceptibility peak, Fig 3c. The RIXS experiment was performed at the ADRESS beamline using the SAXES spectrometer\cite{Strocov2010,Ghiringhelli2006}. Combined, the incident beam and spectrometer gave an overall energy resolution of 240 meV FWHM. At the energy of the Yb $M_5$ edge ($3d_{5/2}\rightarrow 4f$ transition), and 90$^\circ$ scattering angle, the momentum transfer was 1.08 $\AA^{-1}$, which is 125 \%\ of the Brillouin zone width. Due to the well-studied effects of polishing on the optical data \cite{okamura2007}, which mainly effects how abruptly the MIR feature appears as temperature is lowered, we cleaved the rock-hard crystals using a ball-pean hammer and aluminum foil. Shiny facets were smooth, and nearly flat over the beam footprint area (4 x 55 $\mu$m). The effectiveness of this procedure in reducing extrinsic surface contributions can is verified by two methods of monitoring XAS: surface sensitive total electron yield (TEY) collected as sample drain current, and bulk-sensitive total fluorescence yield (TFY) collected via an energy integrating detector give similar results (Fig 1a).

Tender X-ray photons $\hbar\omega_{in}\sim$1518 eV generated from the third harmonic of the ADRESS undulator excite the crystal via a Yb 3$d_{5/2}$$\rightarrow$ 4$f$ dipole transition corresponding to the $M_5$ edge. 
Attempts to fit a Doniach-Sunjic lineshape returned approximately the simpler Lorentzian limit of this function (asymmetry $\beta$$\sim$0.1). This is consistent with a high degree of atomic character of the $d^9f^{14}$ core-hole-containing absorption final state. No $M_4$ edge was observed at high temperature, completely consistent with the Fermi blocking expected from the nearly-full $f$ state manifold inferred from the large Curie moment.

\section{Main results}


\begin{figure}
\begin{center}
\includegraphics[width=3.2in]{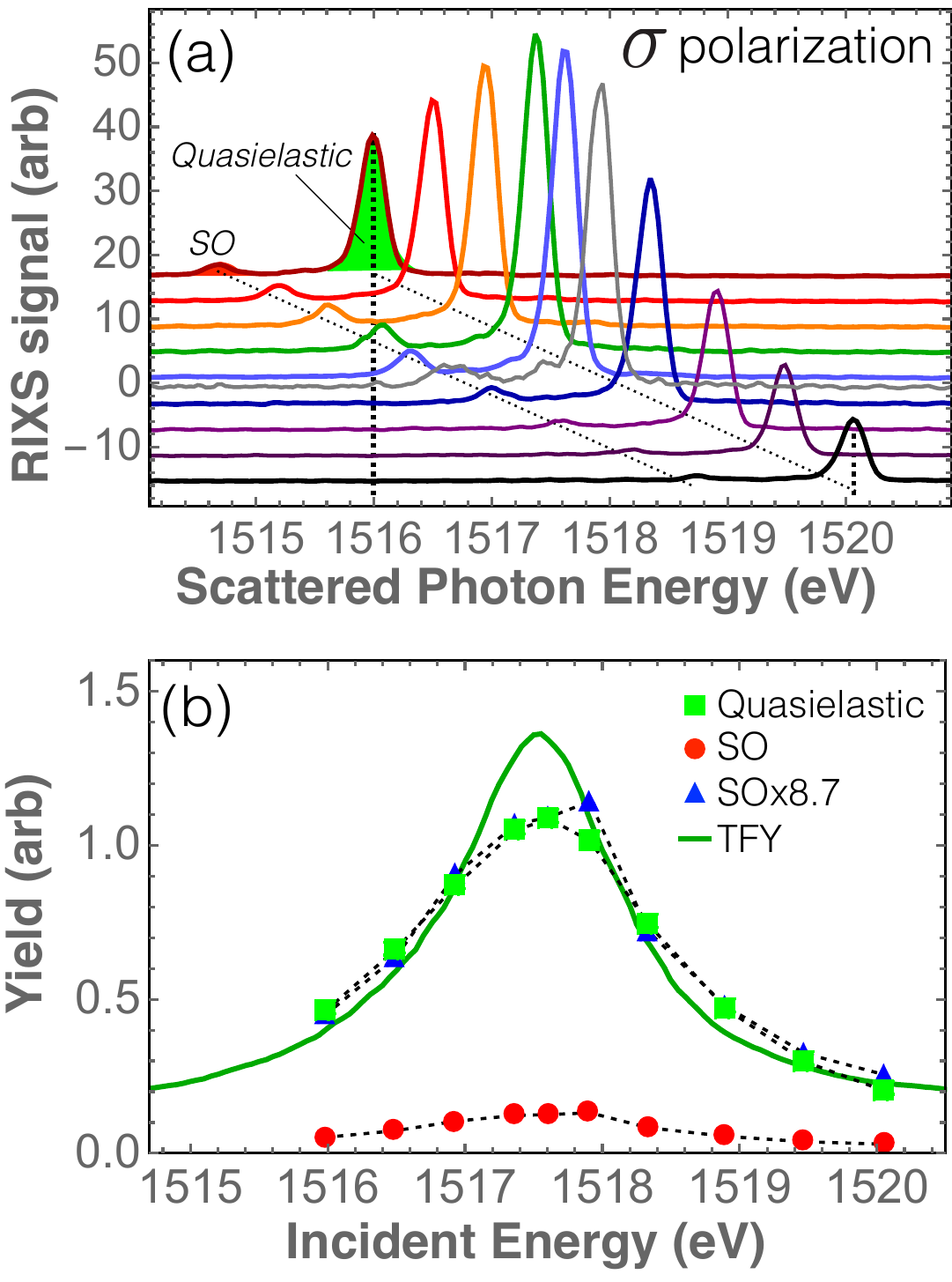}
\caption{(a) RIXS scans at different incident energies through the $M_5$ absorption peak of YbInCu$_4$ at room temperature and $\sigma$ polarization. (b) Dependence of quasielastic and inelastic SO features on incident energy. Marked departure of the integrated RIXS intensity in both cases departs significantly from the Lorentzian line shape of the absorption.}
\label{ }
\end{center}
\end{figure}

RIXS spectra excited near the single resonance peak are represented in Figure 2, and each scan, taken at a different incident energy, shows two main features: the quasielastic peak at zero energy transfer and the smaller feature at 1.4 eV loss arising from the RIXS-active excitation from the filled $j=5/2$ spin-orbit (SO) component of the 4$f$ manifold. This type of excitation is strictly dipole forbidden and therefore inaccessible via infrared spectroscopy. Figure 2b shows the incident energy dependence of the integrated intensity for each feature separately. These RIXS spectra at the $M_5$ edge are exceptionally simple - one inelastic excitation clearly separated from a quasielastic manifold contrasts most high resolution RIXS studies, and reveal that the inelastic signal intensity for this process is not simply proportional to absorption. Below we interpret this deviation as evidence of photon scattering via the indirect RIXS process, wherein core hole interaction plays a significant role. We note that to a very good approximation, it appears the inelastic and elastic signals are proportional to each other, permitting us to focus only on the incident energy which gives the highest RIXS signal as a representative spectrum.

For incident energy tuned to the peak of the absorption, Figure 3a shows a series of 5 minute scans cooling across the valence transition temperature and a clear signature of inelastic excitation appears abruptly below $T_v$, as the system enters the high-$T_K$, moment-screened state. To highlight this electronic response to the transition, we show the integral of the RIXS spectra in the energy band between 200 meV and 400 meV along with the magnetic susceptibility in Fig 3c. A similar abrupt change in the same energy window associated with the valence transition was detected using infrared spectroscopy\cite{Garner2000,Hancock2004b,hancock06a,Zhang2017} (Fig. 3b), and gave the first experimental indication\cite{Garner2000} of the predicted infrared hybridization gap\cite{Coleman1987} in heavy fermion and mixed-valent systems \cite{DeGiorgi2001,Dordevic2001,Chen2016}\footnote{Okamura et al\cite{okamura2007} have shown using infrared microscopy at a synchrotron light source that the abruptness of the appearance of the hybridization gap feature \ybin\ is strongly dependent on the surface history of the sample. Mechanical polishing is a procedure which is commonly used in infrared spectroscopy on complex oxides and other systems, but apparently gives rise to non-intrinsic effects in systems with valence instabilities like \ybin\ \cite{hancock06a} and SmS \cite{batlogg76}}. 

\begin{figure}
\begin{center}
\includegraphics[width=3.2in]{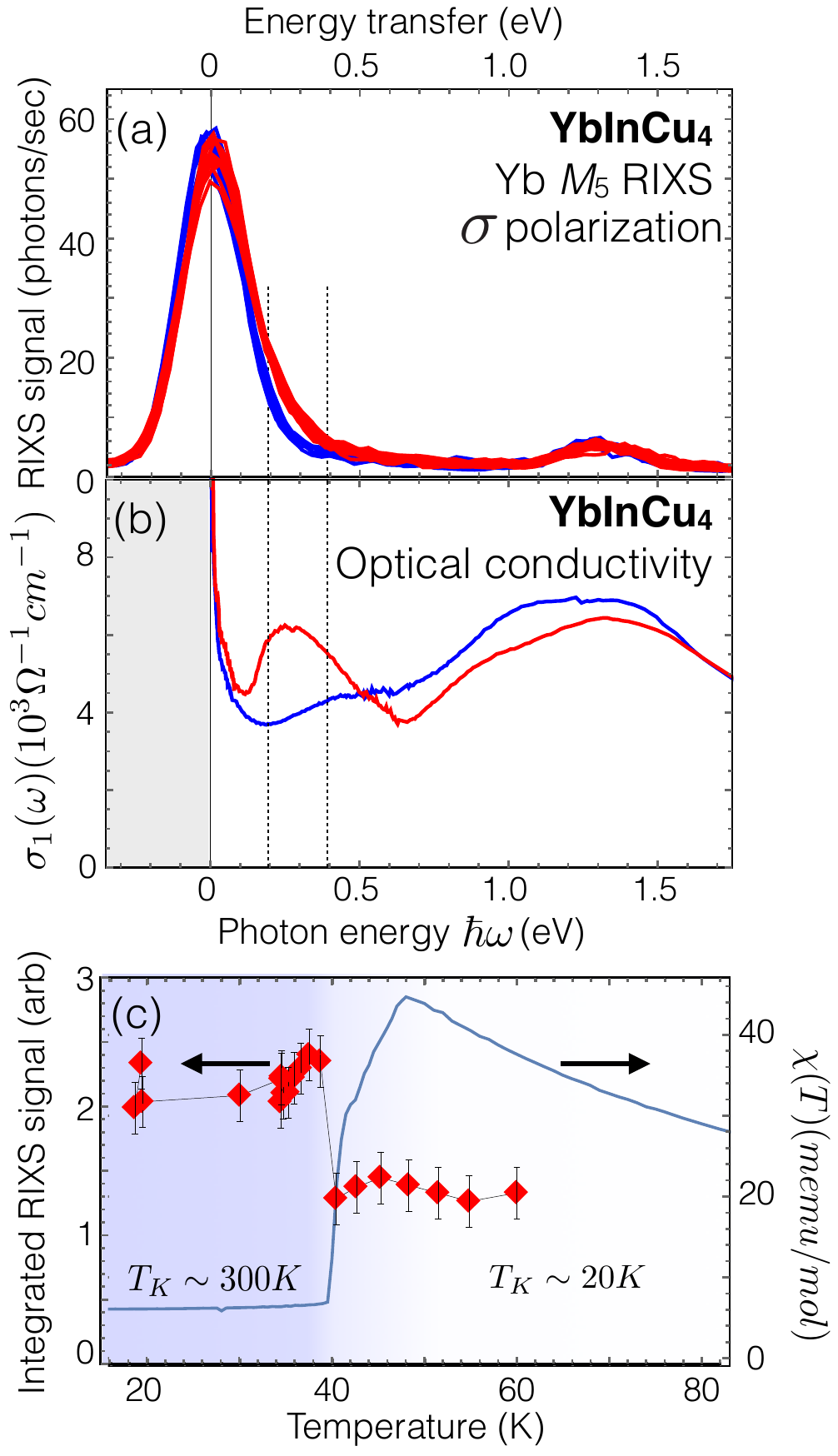}
\caption{(a) 5 minute scans cooling through the valence transition in \ybin, where red lines indicate temperatures below the transition and the blue lines above the transition. (b) optical conductivity at (red) 20K and (blue) 250K from \cite{Hancock2004b}. In (a) and (b),  (c) Intensity integrated between the vertical dotted lines in (a). The step at $T_v$ is clearly displayed. Solid blue trace shows the dc magnetic susceptibility determined through magnetometry.}
\label{ }
\end{center}
\end{figure}

The infrared hybridization gap appears to be a generic feature of heavy fermion physics, and the relation between the energy scale of this feature, the mass enhancement, and the Kondo scale have been well established\cite{Garner2000,DeGiorgi2001,Dordevic2001,Hancock2004b,okamura2007} in the context of infrared measurements. While extremely high-energy resolution is achievable with infrared spectroscopy, one severe limitation to observing the momentum dependence of the hybridization gap is the vanishingly small momentum of infrared photons ($c/\hbar\omega$) in comparison to a typical zone-boundary electronic momentum ($\hbar \pi/a$), as illustrated in Figure 5. Momentum-dependence can be seen indirectly in the context of quasiparticle interference in local tunneling microscopies, as has been demonstrated recently in heavy-fermion \urusi\cite{davis2009,yazdani2010}, or directly for occupied states using angle-resolved photoemission\cite{Wray2015,Kummer2015,Fujimori2016,Patil2016}. Furthermore, optical probes indiscriminantly excite charged excitations, while the RIXS probe is highly species, valence selective with full momentum control, permitting assignment of this feature to the $f$ manifold of the valence-fluctuating Yb ions.

\begin{figure}
\begin{center}
\includegraphics[width=3.2in]{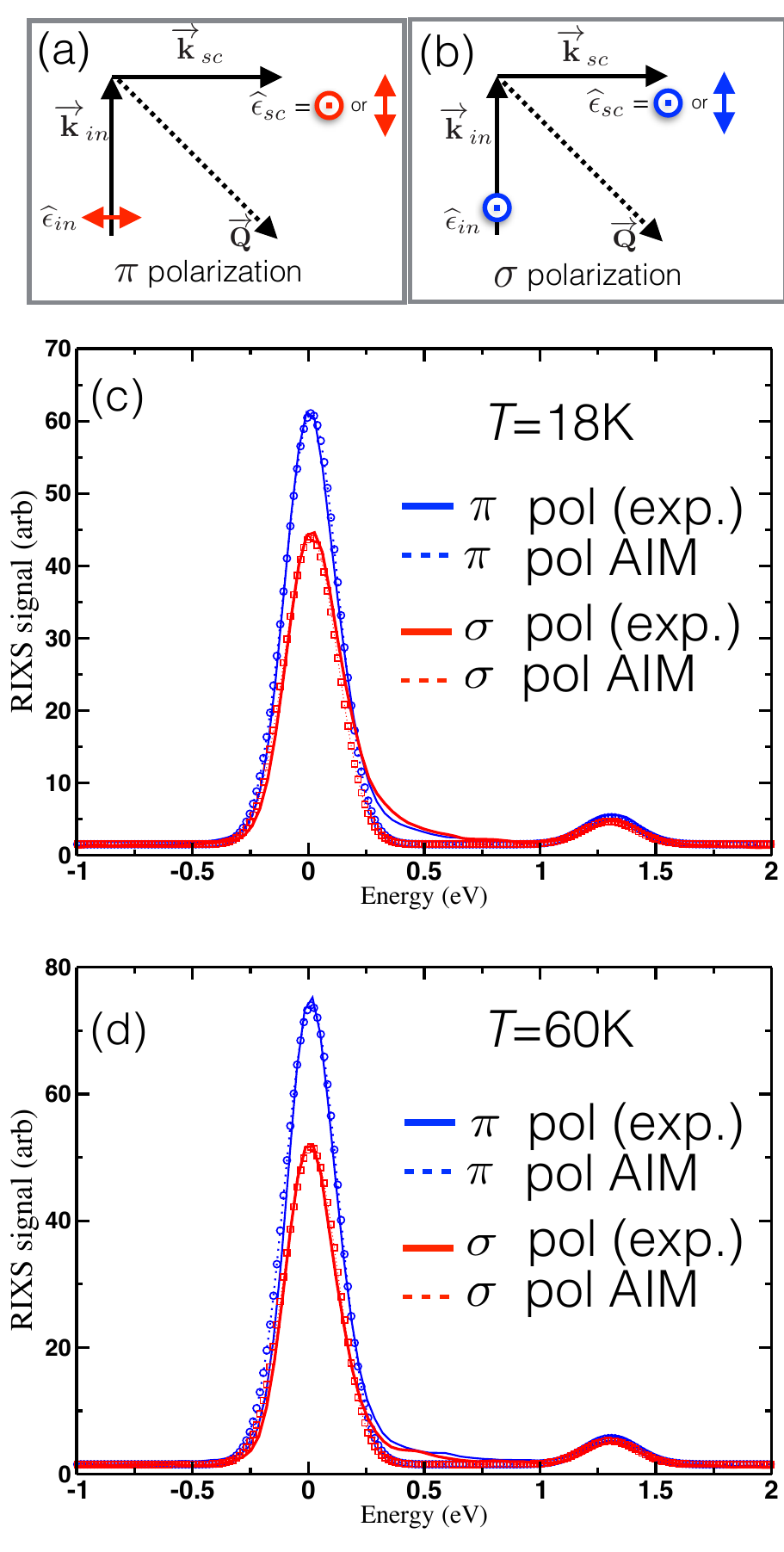}
\caption{(a) and (b) schematic of the scattering geometry for $\sigma$ and $\pi$ polarization conditions. (c) RIXS spectra collected at the absorption peak at 18K$<$42K in the moment-screened phase and (c) at high temperature. Shown in each case is the result of an Anderson impurity model as discussed in the text. The residual RIXS scattering is resultant from indirect RIXS transitions and originates from the lattice system.}
\label{polfig}
\end{center}
\end{figure}

\begin{figure}
\begin{center}
\includegraphics[width=3.5in]{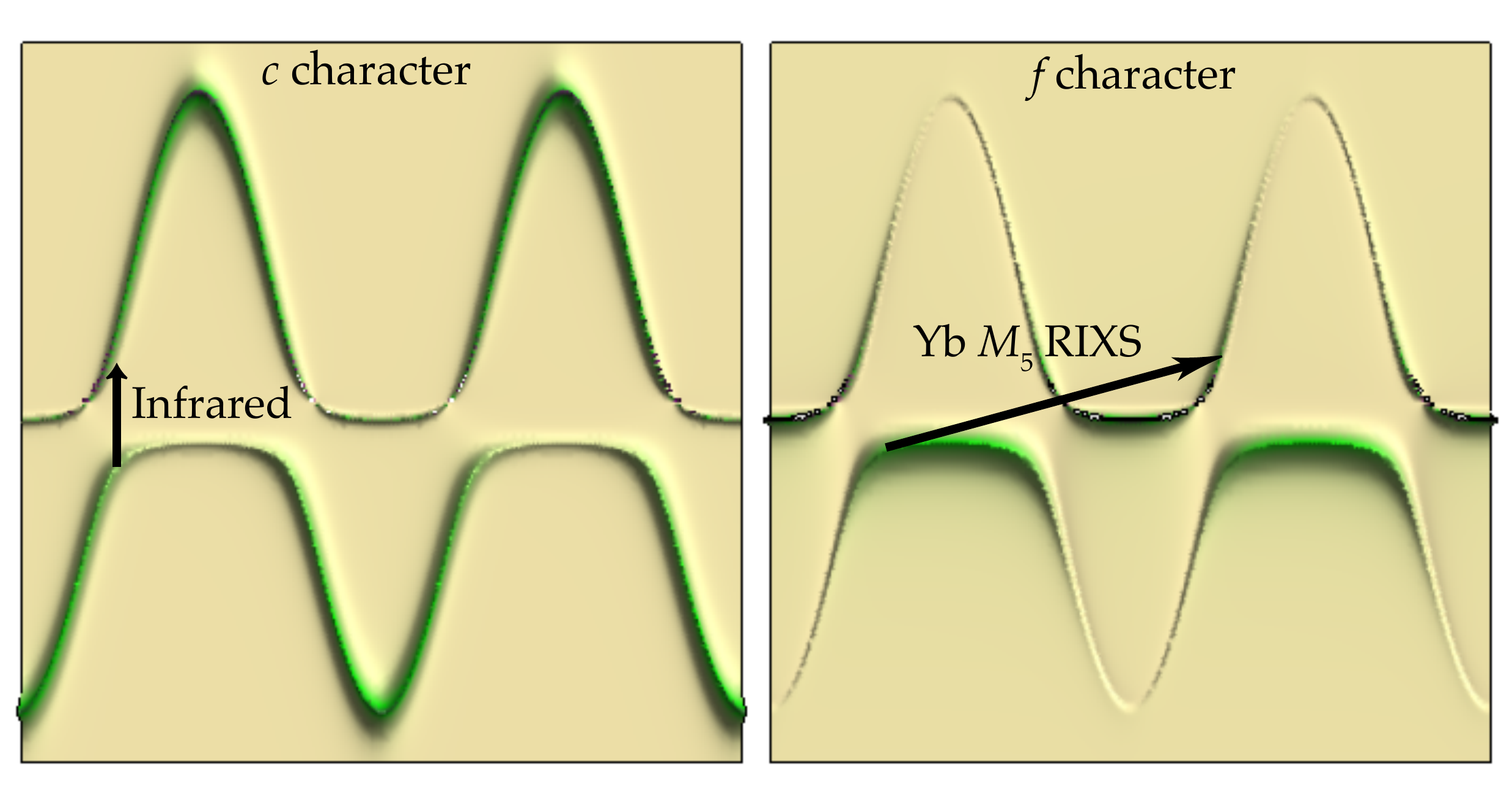}
\caption{Schematic representation of the quasiparticle spectral functions of the periodic Anderson model projected onto states of (left) conduction and (right) localized character. The optical conductivity selectively probes momentum-conserving electron-hole pair excitations, while the RIXS measurements are more heavily influenced by the atomic-like transitions among localized orbitals.}
\label{Fig3}
\end{center}
\end{figure}

\section{Polarization dependence}
To connect the results of our experiments with microscopic quantities of interest in valence-fluctuating materials, we have applied an Anderson impurity model (AIM) appropriate for the free moment state and moment-screened states in two polarization geometries where we have collected RIXS spectra, as described in the appendix. Schematic drawings of the $\pi$ (aka. depolarized) and $\sigma$ (aka polarized) geometries are shown in Figures 4 (a) and (b), respectively. In addition to the RIXS selectivity of atomic species and valence state imbued by the resonance condition, the conservation of angular momentum enables further sensitivities to symmetry which are well-known in Raman scattering \cite{Cardona1983}. The polarization is switched by the elliptically-polarizing undulator in the experiment, and fixed by the matrix elements and appropriate Clebsch-Gordan coefficients in the AIM calculation (see Appendix).

Figure 4c and d respectively show the RIXS spectra in the low-temperature moment-screened phase in $\sigma$ and $\pi$ polarization along with a AIM calculation of the same using the crystal field parameters of Severing et al\cite{Severing1990}. We find that the AIM is capable of describing most of the spectral weight and in particular its polarization dependence, with notable intensity between the quasielastic and SO features which are clearly beyond the AIM. The same exercise in the high-temperature, free moment phase gives remarkable agreement as well. The observation of significant RIXS scattering intensity beyond the AIM which is sensitive to the collective state of this Kondo lattice system is a clear demonstration of the effectiveness of the RIXS probe the low-energy physics of intermediate valence materials constrain theories of correlated electronic behavior. Note that the deviation between the AIM-based theory results and experimental data in the spectral range of order 0.5 eV cannot be due to the crystalline field of lattice excitations, as they reside at the 10meV energy scale. These are most likely governed by the indirect scattering processes during which the electron-electron interactions lead to the creation of the particle-hole pairs which recombine with the excited electron.

\section{Discussion}
The appearance of a RIXS-active inelastic feature simultaneously in the optical conductivity and $M_5$ RIXS is strong evidence that this excitation consists of both itinerant and localized character. In the RIXS probe, strong constraints associated with the atomic transition and weak dipole matrix elements between conduction electrons and the Yb 3$d$ core hole strongly favor $f$-orbital character final states. Conversely, at high temperatures the optical conductivity is primarily governed by the response of the conduction electrons since predominantly localized $f$-electrons have negligible contribution to the optical conductivity. At low temperatures, the conduction and $f$-electrons are strongly hybridized so that the clear distinction between the contributions from the $c$- and $f$-electrons cannot be made. This mixed character is expected based on the predictions of the periodic Anderson model and constitutes the first experimental confirmation of the local character of the incoherent band. 


Although our theory fits based on a single impurity Kondo model work fairly well, the approximations we used naturally miss two possibly important physical aspects of the problem: (i) the effects associated with the onset of the coherent hybridization gap, which is a Kondo lattice phenomenon and (ii) contributions from the indirect RIXS processes. Naturally, the Kondo lattice coherence effects are expected to be pronounced at low temperatures, $T\ll T_K$. The same is also true for the indirect processes since the low-$T$ phase has much higher conductivity compared to its high-$T$ counterpart and, therefore, may lead to a much larger contribution from the particle-hole excitations than in the high-$T$ phase. Since in the indirect RIXS process, the core-hole propagation is the main source of the scattering, possible deviations from the direct RIXS spectra signify the role played by the core hole potential.

Our results demonstrate the sensitivity of RIXS to the hybridization gap in heavy fermion and mixed-valence compounds, and a benchmark for the use of the RIXS technique in providing microscopic information on systems where Coulomb correlated magnetic bands strongly mix with metallic bands. Following the recent work of Kotani \cite{kotani11}, we have shown theoretically that this type of excitation could be observed using Yb $M$ edge RIXS in Yb intermetallics. In future work, the promised ability of RIXS spectrometers to continuously vary the scattering angle will provide the tools needed to deliver detailed information on the momentum-dependence of the hybridization gap in Yb intermediate-valence systems. With subsequent improvements in energy resolution, similar detailed studies in early rare earth species like Ce will also become accessible.

\section{Summary}

In summary, we have demonstrated the first connection between the RIXS technique and the elementary excitation of quasiparticles in the heavy fermion class of materials. We expect this finding is the first in a series of measurements of this type, and will aid to incentivize the development of inelastic X-ray scattering instruments with higher energy resolution. In contrast to optical probes, which mainly probe fast electron-hole pairs with zero relative momentum, the RIXS technique is capable of collecting momentum-dependent information regarding the hybridization gap.



\section{Acknowledgments} We would like to acknowledge valuable conversations with H. Okamura, Z. Schlesinger, and Dirk van der Marel. Support provided by Swiss National Science Foundation, Materials with Novel Electronic Properties (MaNEP) with additional support from the U.S. Department of Energy, Office of Science, Office of Basic Energy Sciences, under Award No. DE-SC0016481. Part of this work was performed at the ADRESS beam line of the Swiss Light Source (SLS) at the Paul Scherrer Institut using the SAXES spectrometer developed jointly by Politecnico di Milano, SLS, and EPFL. M.D. acknowledges the financial support by the National Science Foundation grant NSF-DMR-1506547.

\section{Appendix - Theory}

\subsection{Model parameters}
Consider a trivalent Yb ion in the cubic crystal field, with the $4f$ electrons in the $4f^{13}$ valence configuration. The large spin-orbit coupling splits the $4f$ levels into the $J=7/2$ and $J=5/2$ multiplets. Importantly, the difference between the energies $E_{J=7/2}-E_{J=5/2}=-1.4$ eV, so we can completely ignore the $J=5/2$ multiplet all together in a calculation of the low-energy RIXS spectra including crystal field and hybridization gap features. 

In a cubic crystalline field, the $J=7/2$ multiplet splits into two Kramers doublets $|\Gamma_6\rangle$, $|\Gamma_7\rangle$ and a quartet $|\Gamma_8\rangle$. The corresponding wave functions in $|J,M\rangle$ basis are 
\beg\label{J72wfs}
\begin{split}
&|\Gamma_6 (a=1,2)\rangle=\sqrt{\frac{5}{12}}|\pm 7/2\rangle+\sqrt{\frac{7}{12}}|\mp 1/2\rangle, \\
&|\Gamma_7 (a=1,2)\rangle=\sqrt{\frac{3}{4}}|\pm 5/2\rangle-\sqrt{\frac{1}{4}}|\mp 3/2\rangle, \\
&|\Gamma_8 (a=1,2)\rangle=\sqrt{\frac{7}{12}}|\pm 7/2\rangle-\sqrt{\frac{5}{12}}|\mp 1/2\rangle, \\
&|\Gamma_8 (a=3,4)\rangle=\sqrt{\frac{1}{4}}|\pm 5/2\rangle+\sqrt{\frac{3}{4}}|\mp 3/2\rangle.
\end{split}
\en
In YbInCu$_4$ the ground state multiplet is $\Gamma_8$ \cite{Severing1990} with 
$\Gamma_6$ and $\Gamma_7$ above it with the corresponding energies $E_{\Gamma_6}-E_{\Gamma_8}=3.2$ meV and $E_{\Gamma_7}-E_{\Gamma_8}=3.8$ meV.

\subsection{High-temperature regime, $T>T_v$}
At temperatures above the valence transition temperature $T_v\sim 40$ K the ytterbium $f$-electrons can be approximated as free ions, i.e. the hybridization between the $f$-states and conduction band is considered to be vanishingly small. This view is also supported by the recent RIXS (Yb-$L_3$ edge) observations of the band reconstruction and pronounced changes in the electronic density of states \cite{Jarrige2015} . Within the Kondo volume collapse model, the high temperature regime corresponds to the single ion Kondo temperature of 
$T_{K}^{\textrm{(HT)}}\simeq 25$ K. 

We consider the RIXS process in which the Yb $4f$ hole is excited to the Yb $3d$ intermediate state by the incident photon and a $3d$ hole is de-excited to the Yb $4f$ state by emitting a photon. 
The general expression describing the RIXS intensity reads
\beg\label{RIXS}
\begin{split}
F_{q_1,q_2}(\omega_1,\omega_2)&=\sum\limits_{j,g}\left|\sum\limits_m\frac{\langle{g}|C_{q_1}^{(1)}|m\rangle\langle m|C_{q_2}^{(1)}|f\rangle}{E_g+\omega_1-E_m+i\gamma}\right|^2\\&\times G(\omega_1-\omega_2-E_{\textrm{fin}}+E_{g})
\end{split}
\en
where $\omega_1$ is an energy of an incident photon with polarization $q_1=-1,0,1$, $\omega_2$ is an energy of emitted photon with polarization $q_2$, $\gamma$ is a lifetime broadening of the intermediate state, $C_{q}^{(1)}$ is the spherical tensor operator 
corresponding to the electric dipole transition due to a linear coupling between the electron's momentum and electromagmetic vector potential, and $G(\epsilon)$ is some function which accounts for the experimental resolution and the states $|g\rangle,|m\rangle,|f\rangle$ are
initial (i.e. ground), intermediate and final states correspondingly.

For the case when the single photon energy is tuned to the resonance energy, which is the energy difference between the single intermediate state and the final $f$-hole state:
\beg\label{Resonance}
E_m=E_g+\omega_1,
\en
the RIXS spectrum is governed by the following expression
\begin{widetext}
\beg\label{RIXSResonance}
\begin{split}
F_{q_1,q_2}(\omega_1,\omega_2)&=\frac{1}{\gamma^2}\sum\limits_{i\alpha\alpha'}\left|\sum\limits_{j_z}
\langle 4f;\Gamma_8(\alpha')|C_{q_1}^{(1)}|3d;j_z\rangle
\langle3d;j_z|C_{q_2}^{(1)}|4f;\Gamma_i(\alpha)\rangle\right|^2
G[\omega_1-\omega_2-E(\Gamma_i)+E(\Gamma_8)].
\end{split}
\en
\end{widetext}
In this expression it is assumed that the $f$-electron is in $\Gamma_8$ multiplet. Summation is performed over the initial ($\alpha$) and final ($\alpha'$) components of the $4f$ multiplet. To compute the matrix elements of the spherical operator, we write
\beg\label{Matrix}
\begin{split}
\langle 3d;j_z|C_{q}^{(1)}|4f;\Gamma_i(\alpha)\rangle&=\sum\limits_{m_z=-7/2}^{7/2}\langle 5/2~j_z~1~q|7/2~m_z\rangle\\&\times\langle7/2;m_z|4f;\Gamma_i(\alpha)\rangle,
\end{split}
\en
where $i=6,7,8$.
Here $\langle 5/2~j_z~1~q|7/2~m_z\rangle$ are the Clebsh-Gordan coefficients and the matrix elements 
$\langle7/2;m_z|4f;\Gamma(\alpha)\rangle$
are computed using the wave-functions (\ref{J72wfs}). Note that the Clebsh-Gordan coefficient to be nonzero the following condition
\beg
j_z+q=m_z
\en
must be met. The results of the calculation for various polarizations are summarized in Fig. \ref{polfig}.

\subsection{Low-temperature regime, $T<T_v$}
At low temperatures, the ground state of the system is the singlet state corresponding to the Kondo singlet bound state. Since the magnetic correlations in YbInCu$_4$ seem to be negligible, we follow the discussion in Ref. \onlinecite{kotani11} and consider the single-impurity Anderson model (SIAM):
\beg\label{SIAM}
\begin{split}
\hat{H}&=\sum\limits_{\nu=-J}^J\hat{h}_\nu+\hat{H}_{int}, \\
\hat{h}_\nu&=\epsilon_{f}\hat{f}_\nu^\dagger\hat{f}_\nu+\sum\limits_{k}\varepsilon_k\hat{c}_{k\nu}^\dagger\hat{c}_{k\nu}+V\sum\limits_{k}(\hat{c}_{k\nu}^\dagger\hat{f}_\nu+\textrm{h.c.})\\&+U_{ff}\sum\limits_{\nu'\not=\nu}\hat{f}_\nu^\dagger\hat{f}_\nu\hat{f}_{\nu'}^\dagger\hat{f}_{\nu'}
\end{split}
\en
where the $f$ states are represented in $JJ$ coupling scheme and $\nu=-J,...,J$ with $J=7/2$, $\hat{H}_{\textrm{int}}$ accounts for an intermediate state with the $d$-hole and filled $f$-shell, and $U_{\textrm{ff}}>0$ is the short-range interaction between the $f$-electrons. 
At zero temperature the ground state is the superposition of $4f^{14}$ and $4f^{13}c$ electronic configurations. 

In the limit $U_{\textrm{ff}}\to\infty$ one can employ the large degeneracy of the $f$-electron multiplet to describe the ground state properties qualitatively. For the calculation of the RIXS spectra, however, one can employ the variational principle first discussed by Varma and Yafet \cite{Varma1976} which for the ground state energy yields \cite{kotani11}:
\beg\label{Eg}
E_g=\epsilon_f-k_BT_K+\sum\limits_{k\leq k_F}\epsilon_k.
\en
Here $T_K$ is the single ion Kondo temperature, $\epsilon_f\approx -1$ eV and the momentum summation is performed over the occupied states in the conduction band ($k_F$ is the Fermi momentum). In the case of the inelastic scattering, similarly to the above one obtains the energy of the final state
\beg\label{Ef}
E_f=E_g+k_BT_K-k_B\tilde{T}_K,
\en
where $\tilde{T}_K$ is found by solving the following system of equations
\beg\label{EqFinal}
\begin{split}
k_B\tilde{T}_K-\epsilon_f+V\sum\limits_{k}\theta(k-k_F)a_k&=0, \\
(k_B\tilde{T}_K+\epsilon_k-\varepsilon_f)a_k+V&=0,
\end{split}
\en
Here, the momentum summation is limited to the width of the conduction band, $k\in[-W/2,W/2]$ and the coefficients $a_k$ depend on the value of $k_BT_K$. For the realistic values of the model parameters, $\tilde{T}_K\ll T_K$. 
Our theoretical plot for the intensity and two different polarizations are shown in Fig. \ref{polfig}. 
Note, that our results show that even within the resolution $R\sim 240$ meV, the spectroscopic features of the Kondo resonance are visible thus providing yet another insight into the physics of the isostructural valence transition. 

%
%

\bibliography{kondo,rixs,library}

\begin{thebibliography}{33}
\expandafter\ifx\csname natexlab\endcsname\relax\def\natexlab#1{#1}\fi
\expandafter\ifx\csname bibnamefont\endcsname\relax
  \def\bibnamefont#1{#1}\fi
\expandafter\ifx\csname bibfnamefont\endcsname\relax
  \def\bibfnamefont#1{#1}\fi
\expandafter\ifx\csname citenamefont\endcsname\relax
  \def\citenamefont#1{#1}\fi
\expandafter\ifx\csname url\endcsname\relax
  \def\url#1{\texttt{#1}}\fi
\expandafter\ifx\csname urlprefix\endcsname\relax\def\urlprefix{URL }\fi
\providecommand{\bibinfo}[2]{#2}
\providecommand{\eprint}[2][]{\url{#2}}

\bibitem[{\citenamefont{Allen and Martin}(1982)}]{Allen1982}
\bibinfo{author}{\bibfnamefont{J.~W.} \bibnamefont{Allen}} \bibnamefont{and}
  \bibinfo{author}{\bibfnamefont{R.~M.} \bibnamefont{Martin}},
  \bibinfo{journal}{Physical Review Letters} \textbf{\bibinfo{volume}{49}},
  \bibinfo{pages}{1106} (\bibinfo{year}{1982}), ISSN \bibinfo{issn}{0031-9007},
  \urlprefix\url{http://link.aps.org/doi/10.1103/PhysRevLett.49.1106}.

\bibitem[{\citenamefont{Lanat{\`{a}} et~al.}(2013)\citenamefont{Lanat{\`{a}},
  Yao, Wang, Ho, Schmalian, Haule, and Kotliar}}]{Lanata2013}
\bibinfo{author}{\bibfnamefont{N.}~\bibnamefont{Lanat{\`{a}}}},
  \bibinfo{author}{\bibfnamefont{Y.-X.} \bibnamefont{Yao}},
  \bibinfo{author}{\bibfnamefont{C.-Z.} \bibnamefont{Wang}},
  \bibinfo{author}{\bibfnamefont{K.-M.} \bibnamefont{Ho}},
  \bibinfo{author}{\bibfnamefont{J.}~\bibnamefont{Schmalian}},
  \bibinfo{author}{\bibfnamefont{K.}~\bibnamefont{Haule}}, \bibnamefont{and}
  \bibinfo{author}{\bibfnamefont{G.}~\bibnamefont{Kotliar}},
  \bibinfo{journal}{Physical Review Letters} \textbf{\bibinfo{volume}{111}},
  \bibinfo{pages}{196801} (\bibinfo{year}{2013}), ISSN
  \bibinfo{issn}{0031-9007},
  \urlprefix\url{http://link.aps.org/doi/10.1103/PhysRevLett.111.196801}.

\bibitem[{\citenamefont{Dzero et~al.}(2000)\citenamefont{Dzero, Gor'kov, and
  Zvezdin}}]{Dzero2000}
\bibinfo{author}{\bibfnamefont{M.~O.} \bibnamefont{Dzero}},
  \bibinfo{author}{\bibfnamefont{L.~P.} \bibnamefont{Gor'kov}},
  \bibnamefont{and} \bibinfo{author}{\bibfnamefont{A.~K.}
  \bibnamefont{Zvezdin}}, \bibinfo{journal}{Journal of Physics: Condensed
  Matter} \textbf{\bibinfo{volume}{12}}, \bibinfo{pages}{L711}
  (\bibinfo{year}{2000}), ISSN \bibinfo{issn}{0953-8984},
  \urlprefix\url{http://stacks.iop.org/0953-8984/12/i=47/a=102?key=crossref.1c4b50456da691e1a0083a7dbf028bc6}.

\bibitem[{\citenamefont{Howald et~al.}(2015)\citenamefont{Howald, Stilp,
  de~R{\'{e}}otier, Yaouanc, Raymond, Piamonteze, Lapertot, Baines, and
  Keller}}]{Howald2015}
\bibinfo{author}{\bibfnamefont{L.}~\bibnamefont{Howald}},
  \bibinfo{author}{\bibfnamefont{E.}~\bibnamefont{Stilp}},
  \bibinfo{author}{\bibfnamefont{P.~D.} \bibnamefont{de~R{\'{e}}otier}},
  \bibinfo{author}{\bibfnamefont{A.}~\bibnamefont{Yaouanc}},
  \bibinfo{author}{\bibfnamefont{S.}~\bibnamefont{Raymond}},
  \bibinfo{author}{\bibfnamefont{C.}~\bibnamefont{Piamonteze}},
  \bibinfo{author}{\bibfnamefont{G.}~\bibnamefont{Lapertot}},
  \bibinfo{author}{\bibfnamefont{C.}~\bibnamefont{Baines}}, \bibnamefont{and}
  \bibinfo{author}{\bibfnamefont{H.}~\bibnamefont{Keller}},
  \textbf{\bibinfo{volume}{5}}, \bibinfo{pages}{12528} (\bibinfo{year}{2015}),
  \urlprefix\url{http://dx.doi.org/10.1038/srep12528 http://10.0.4.14/srep12528
  https://www.nature.com/articles/srep12528{\#}supplementary-information}.

\bibitem[{\citenamefont{Hancock et~al.}(2004)\citenamefont{Hancock, McKnew,
  Schlesinger, Sarrao, and Fisk}}]{Hancock2004b}
\bibinfo{author}{\bibfnamefont{J.}~\bibnamefont{Hancock}},
  \bibinfo{author}{\bibfnamefont{T.}~\bibnamefont{McKnew}},
  \bibinfo{author}{\bibfnamefont{Z.}~\bibnamefont{Schlesinger}},
  \bibinfo{author}{\bibfnamefont{J.}~\bibnamefont{Sarrao}}, \bibnamefont{and}
  \bibinfo{author}{\bibfnamefont{Z.}~\bibnamefont{Fisk}},
  \bibinfo{journal}{Physical Review Letters} \textbf{\bibinfo{volume}{92}},
  \bibinfo{pages}{186405} (\bibinfo{year}{2004}), ISSN
  \bibinfo{issn}{0031-9007},
  \urlprefix\url{http://link.aps.org/doi/10.1103/PhysRevLett.92.186405}.

\bibitem[{\citenamefont{Hancock et~al.}(2006)\citenamefont{Hancock, McKnew,
  Schlesinger, Sarrao, and Fisk}}]{hancock06a}
\bibinfo{author}{\bibfnamefont{J.~N.} \bibnamefont{Hancock}},
  \bibinfo{author}{\bibfnamefont{T.}~\bibnamefont{McKnew}},
  \bibinfo{author}{\bibfnamefont{Z.}~\bibnamefont{Schlesinger}},
  \bibinfo{author}{\bibfnamefont{J.~L.} \bibnamefont{Sarrao}},
  \bibnamefont{and} \bibinfo{author}{\bibfnamefont{Z.}~\bibnamefont{Fisk}},
  \bibinfo{journal}{Phys. Rev. B} \textbf{\bibinfo{volume}{73}},
  \bibinfo{pages}{125119} (\bibinfo{year}{2006}).

\bibitem[{\citenamefont{Cornelius et~al.}(1997)\citenamefont{Cornelius,
  Lawrence, Sarrao, Fisk, Hundley, Kwei, Thompson, Booth, and
  Bridges}}]{corn98}
\bibinfo{author}{\bibfnamefont{A.~L.} \bibnamefont{Cornelius}},
  \bibinfo{author}{\bibfnamefont{J.~M.} \bibnamefont{Lawrence}},
  \bibinfo{author}{\bibfnamefont{J.~L.} \bibnamefont{Sarrao}},
  \bibinfo{author}{\bibfnamefont{Z.}~\bibnamefont{Fisk}},
  \bibinfo{author}{\bibfnamefont{M.~F.} \bibnamefont{Hundley}},
  \bibinfo{author}{\bibfnamefont{G.~H.} \bibnamefont{Kwei}},
  \bibinfo{author}{\bibfnamefont{J.~D.} \bibnamefont{Thompson}},
  \bibinfo{author}{\bibfnamefont{C.~H.} \bibnamefont{Booth}}, \bibnamefont{and}
  \bibinfo{author}{\bibfnamefont{F.}~\bibnamefont{Bridges}},
  \bibinfo{journal}{Phys. Rev. B} \textbf{\bibinfo{volume}{56}},
  \bibinfo{pages}{7993} (\bibinfo{year}{1997}).

\bibitem[{\citenamefont{Jarrige et~al.}(2015)\citenamefont{Jarrige, Kotani,
  Yamaoka, Tsujii, Ishii, Upton, Casa, Kim, Gog, and Hancock}}]{Jarrige2015}
\bibinfo{author}{\bibfnamefont{I.}~\bibnamefont{Jarrige}},
  \bibinfo{author}{\bibfnamefont{A.}~\bibnamefont{Kotani}},
  \bibinfo{author}{\bibfnamefont{H.}~\bibnamefont{Yamaoka}},
  \bibinfo{author}{\bibfnamefont{N.}~\bibnamefont{Tsujii}},
  \bibinfo{author}{\bibfnamefont{K.}~\bibnamefont{Ishii}},
  \bibinfo{author}{\bibfnamefont{M.}~\bibnamefont{Upton}},
  \bibinfo{author}{\bibfnamefont{D.}~\bibnamefont{Casa}},
  \bibinfo{author}{\bibfnamefont{J.}~\bibnamefont{Kim}},
  \bibinfo{author}{\bibfnamefont{T.}~\bibnamefont{Gog}}, \bibnamefont{and}
  \bibinfo{author}{\bibfnamefont{J.}~\bibnamefont{Hancock}},
  \bibinfo{journal}{Physical Review Letters} \textbf{\bibinfo{volume}{114}},
  \bibinfo{pages}{126401} (\bibinfo{year}{2015}), ISSN
  \bibinfo{issn}{0031-9007},
  \urlprefix\url{http://link.aps.org/doi/10.1103/PhysRevLett.114.126401}.

\bibitem[{\citenamefont{Figueroa et~al.}(1998)\citenamefont{Figueroa, Lawrence,
  Sarrao, Fisk, Hundley, and Thompson}}]{Figueroa1998}
\bibinfo{author}{\bibfnamefont{E.}~\bibnamefont{Figueroa}},
  \bibinfo{author}{\bibfnamefont{J.}~\bibnamefont{Lawrence}},
  \bibinfo{author}{\bibfnamefont{J.}~\bibnamefont{Sarrao}},
  \bibinfo{author}{\bibfnamefont{Z.}~\bibnamefont{Fisk}},
  \bibinfo{author}{\bibfnamefont{M.}~\bibnamefont{Hundley}}, \bibnamefont{and}
  \bibinfo{author}{\bibfnamefont{J.}~\bibnamefont{Thompson}},
  \bibinfo{journal}{Solid State Communications} \textbf{\bibinfo{volume}{106}},
  \bibinfo{pages}{347} (\bibinfo{year}{1998}), ISSN \bibinfo{issn}{00381098},
  \urlprefix\url{http://www.sciencedirect.com/science/article/pii/S0038109898000428}.

\bibitem[{\citenamefont{Sarrao et~al.}(1998)\citenamefont{Sarrao, Ramirez,
  Darling, Freibert, Migliori, Immer, Fisk, and Uwatoko}}]{Sarrao1998}
\bibinfo{author}{\bibfnamefont{J.}~\bibnamefont{Sarrao}},
  \bibinfo{author}{\bibfnamefont{A.}~\bibnamefont{Ramirez}},
  \bibinfo{author}{\bibfnamefont{T.}~\bibnamefont{Darling}},
  \bibinfo{author}{\bibfnamefont{F.}~\bibnamefont{Freibert}},
  \bibinfo{author}{\bibfnamefont{A.}~\bibnamefont{Migliori}},
  \bibinfo{author}{\bibfnamefont{C.}~\bibnamefont{Immer}},
  \bibinfo{author}{\bibfnamefont{Z.}~\bibnamefont{Fisk}}, \bibnamefont{and}
  \bibinfo{author}{\bibfnamefont{Y.}~\bibnamefont{Uwatoko}},
  \bibinfo{journal}{Physical Review B} \textbf{\bibinfo{volume}{58}},
  \bibinfo{pages}{409} (\bibinfo{year}{1998}), ISSN \bibinfo{issn}{0163-1829}.

\bibitem[{\citenamefont{Amorese et~al.}(2016)\citenamefont{Amorese, Dellea,
  Fanciulli, Seiro, Geibel, Krellner, Makarova, Braicovich, Ghiringhelli,
  Vyalikh et~al.}}]{Amorese2016}
\bibinfo{author}{\bibfnamefont{A.}~\bibnamefont{Amorese}},
  \bibinfo{author}{\bibfnamefont{G.}~\bibnamefont{Dellea}},
  \bibinfo{author}{\bibfnamefont{M.}~\bibnamefont{Fanciulli}},
  \bibinfo{author}{\bibfnamefont{S.}~\bibnamefont{Seiro}},
  \bibinfo{author}{\bibfnamefont{C.}~\bibnamefont{Geibel}},
  \bibinfo{author}{\bibfnamefont{C.}~\bibnamefont{Krellner}},
  \bibinfo{author}{\bibfnamefont{I.~P.} \bibnamefont{Makarova}},
  \bibinfo{author}{\bibfnamefont{L.}~\bibnamefont{Braicovich}},
  \bibinfo{author}{\bibfnamefont{G.}~\bibnamefont{Ghiringhelli}},
  \bibinfo{author}{\bibfnamefont{D.~V.} \bibnamefont{Vyalikh}},
  \bibnamefont{et~al.}, \bibinfo{journal}{Physical Review B}
  \textbf{\bibinfo{volume}{93}}, \bibinfo{pages}{165134}
  (\bibinfo{year}{2016}), ISSN \bibinfo{issn}{2469-9950},
  \urlprefix\url{https://link.aps.org/doi/10.1103/PhysRevB.93.165134}.

\bibitem[{\citenamefont{Garner et~al.}(2000)\citenamefont{Garner, Hancock,
  Rodriguez, Schlesinger, Bucher, Fisk, and Sarrao}}]{Garner2000}
\bibinfo{author}{\bibfnamefont{S.}~\bibnamefont{Garner}},
  \bibinfo{author}{\bibfnamefont{J.}~\bibnamefont{Hancock}},
  \bibinfo{author}{\bibfnamefont{Y.}~\bibnamefont{Rodriguez}},
  \bibinfo{author}{\bibfnamefont{Z.}~\bibnamefont{Schlesinger}},
  \bibinfo{author}{\bibfnamefont{B.}~\bibnamefont{Bucher}},
  \bibinfo{author}{\bibfnamefont{Z.}~\bibnamefont{Fisk}}, \bibnamefont{and}
  \bibinfo{author}{\bibfnamefont{J.}~\bibnamefont{Sarrao}},
  \bibinfo{journal}{Physical Review B} \textbf{\bibinfo{volume}{62}},
  \bibinfo{pages}{R4778} (\bibinfo{year}{2000}), ISSN
  \bibinfo{issn}{0163-1829},
  \urlprefix\url{http://link.aps.org/doi/10.1103/PhysRevB.62.R4778}.

\bibitem[{\citenamefont{Degiorgi et~al.}(2001)\citenamefont{Degiorgi, Anders,
  and Gr{\"{u}}ner}}]{DeGiorgi2001}
\bibinfo{author}{\bibfnamefont{L.}~\bibnamefont{Degiorgi}},
  \bibinfo{author}{\bibfnamefont{F.}~\bibnamefont{Anders}}, \bibnamefont{and}
  \bibinfo{author}{\bibfnamefont{G.}~\bibnamefont{Gr{\"{u}}ner}},
  \bibinfo{journal}{The European Physical Journal B}
  \textbf{\bibinfo{volume}{19}}, \bibinfo{pages}{167} (\bibinfo{year}{2001}),
  ISSN \bibinfo{issn}{1434-6028},
  \urlprefix\url{http://link.springer.com/10.1007/s100510170324}.

\bibitem[{\citenamefont{Dordevic et~al.}(2001)\citenamefont{Dordevic, Basov,
  Dilley, Bauer, and Maple}}]{Dordevic2001}
\bibinfo{author}{\bibfnamefont{S.~V.} \bibnamefont{Dordevic}},
  \bibinfo{author}{\bibfnamefont{D.~N.} \bibnamefont{Basov}},
  \bibinfo{author}{\bibfnamefont{N.~R.} \bibnamefont{Dilley}},
  \bibinfo{author}{\bibfnamefont{E.~D.} \bibnamefont{Bauer}}, \bibnamefont{and}
  \bibinfo{author}{\bibfnamefont{M.~B.} \bibnamefont{Maple}},
  \bibinfo{journal}{Physical Review Letters} \textbf{\bibinfo{volume}{86}},
  \bibinfo{pages}{684} (\bibinfo{year}{2001}), ISSN \bibinfo{issn}{0031-9007},
  \urlprefix\url{http://link.aps.org/doi/10.1103/PhysRevLett.86.684}.

\bibitem[{\citenamefont{Okamura et~al.}(2007)\citenamefont{Okamura, Michizawa,
  Nanba, and Ebihara}}]{okamura2007}
\bibinfo{author}{\bibfnamefont{H.}~\bibnamefont{Okamura}},
  \bibinfo{author}{\bibfnamefont{T.}~\bibnamefont{Michizawa}},
  \bibinfo{author}{\bibfnamefont{T.}~\bibnamefont{Nanba}}, \bibnamefont{and}
  \bibinfo{author}{\bibfnamefont{T.}~\bibnamefont{Ebihara}},
  \bibinfo{journal}{Phys. Rev. B} \textbf{\bibinfo{volume}{75}},
  \bibinfo{pages}{041101} (\bibinfo{year}{2007}).

\bibitem[{\citenamefont{Hancock}(2004)}]{Hancock2004a}
\bibinfo{author}{\bibfnamefont{J.}~\bibnamefont{Hancock}}, Ph.D. thesis,
  \bibinfo{school}{University of California} (\bibinfo{year}{2004}).

\bibitem[{\citenamefont{Sarrao}(1999)}]{Sarrao1999}
\bibinfo{author}{\bibfnamefont{J.}~\bibnamefont{Sarrao}},
  \bibinfo{journal}{Physica B: Condensed Matter}
  \textbf{\bibinfo{volume}{259-261}}, \bibinfo{pages}{128}
  (\bibinfo{year}{1999}), ISSN \bibinfo{issn}{09214526}.

\bibitem[{\citenamefont{Strocov et~al.}(2010)\citenamefont{Strocov, Schmitt,
  Flechsig, Schmidt, Imhof, Chen, Raabe, Betemps, Zimoch, Krempasky
  et~al.}}]{Strocov2010}
\bibinfo{author}{\bibfnamefont{V.~N.} \bibnamefont{Strocov}},
  \bibinfo{author}{\bibfnamefont{T.}~\bibnamefont{Schmitt}},
  \bibinfo{author}{\bibfnamefont{U.}~\bibnamefont{Flechsig}},
  \bibinfo{author}{\bibfnamefont{T.}~\bibnamefont{Schmidt}},
  \bibinfo{author}{\bibfnamefont{A.}~\bibnamefont{Imhof}},
  \bibinfo{author}{\bibfnamefont{Q.}~\bibnamefont{Chen}},
  \bibinfo{author}{\bibfnamefont{J.}~\bibnamefont{Raabe}},
  \bibinfo{author}{\bibfnamefont{R.}~\bibnamefont{Betemps}},
  \bibinfo{author}{\bibfnamefont{D.}~\bibnamefont{Zimoch}},
  \bibinfo{author}{\bibfnamefont{J.}~\bibnamefont{Krempasky}},
  \bibnamefont{et~al.}, \bibinfo{journal}{Journal of Synchrotron Radiation}
  \textbf{\bibinfo{volume}{17}}, \bibinfo{pages}{631} (\bibinfo{year}{2010}),
  ISSN \bibinfo{issn}{0909-0495},
  \urlprefix\url{http://www.ncbi.nlm.nih.gov/pubmed/20724785
  http://www.pubmedcentral.nih.gov/articlerender.fcgi?artid=PMC2927903
  http://scripts.iucr.org/cgi-bin/paper?S0909049510019862}.

\bibitem[{\citenamefont{Ghiringhelli et~al.}(2006)\citenamefont{Ghiringhelli,
  Piazzalunga, Dallera, Trezzi, Braicovich, Schmitt, Strocov, Betemps, Patthey,
  Wang et~al.}}]{Ghiringhelli2006}
\bibinfo{author}{\bibfnamefont{G.}~\bibnamefont{Ghiringhelli}},
  \bibinfo{author}{\bibfnamefont{A.}~\bibnamefont{Piazzalunga}},
  \bibinfo{author}{\bibfnamefont{C.}~\bibnamefont{Dallera}},
  \bibinfo{author}{\bibfnamefont{G.}~\bibnamefont{Trezzi}},
  \bibinfo{author}{\bibfnamefont{L.}~\bibnamefont{Braicovich}},
  \bibinfo{author}{\bibfnamefont{T.}~\bibnamefont{Schmitt}},
  \bibinfo{author}{\bibfnamefont{V.~N.} \bibnamefont{Strocov}},
  \bibinfo{author}{\bibfnamefont{R.}~\bibnamefont{Betemps}},
  \bibinfo{author}{\bibfnamefont{L.}~\bibnamefont{Patthey}},
  \bibinfo{author}{\bibfnamefont{X.}~\bibnamefont{Wang}}, \bibnamefont{et~al.},
  \bibinfo{journal}{Review of Scientific Instruments}
  \textbf{\bibinfo{volume}{77}}, \bibinfo{pages}{113108}
  (\bibinfo{year}{2006}), ISSN \bibinfo{issn}{0034-6748},
  \urlprefix\url{http://aip.scitation.org/doi/10.1063/1.2372731}.

\bibitem[{\citenamefont{Zhang et~al.}(2017)\citenamefont{Zhang, Chen, Dong, and
  Wang}}]{Zhang2017}
\bibinfo{author}{\bibfnamefont{M.~Y.} \bibnamefont{Zhang}},
  \bibinfo{author}{\bibfnamefont{R.~Y.} \bibnamefont{Chen}},
  \bibinfo{author}{\bibfnamefont{T.}~\bibnamefont{Dong}}, \bibnamefont{and}
  \bibinfo{author}{\bibfnamefont{N.~L.} \bibnamefont{Wang}},
  \bibinfo{journal}{Physical Review B} \textbf{\bibinfo{volume}{95}},
  \bibinfo{pages}{165104} (\bibinfo{year}{2017}), ISSN
  \bibinfo{issn}{2469-9950},
  \urlprefix\url{http://link.aps.org/doi/10.1103/PhysRevB.95.165104}.

\bibitem[{\citenamefont{Coleman}(1987)}]{Coleman1987}
\bibinfo{author}{\bibfnamefont{P.}~\bibnamefont{Coleman}},
  \bibinfo{journal}{Physical Review Letters} \textbf{\bibinfo{volume}{59}},
  \bibinfo{pages}{1026} (\bibinfo{year}{1987}), ISSN \bibinfo{issn}{0031-9007},
  \urlprefix\url{http://link.aps.org/doi/10.1103/PhysRevLett.59.1026}.

\bibitem[{\citenamefont{Chen and Wang}(2016)}]{Chen2016}
\bibinfo{author}{\bibfnamefont{R.~Y.} \bibnamefont{Chen}} \bibnamefont{and}
  \bibinfo{author}{\bibfnamefont{N.~L.} \bibnamefont{Wang}},
  \bibinfo{journal}{Reports on Progress in Physics}
  \textbf{\bibinfo{volume}{79}}, \bibinfo{pages}{064502}
  (\bibinfo{year}{2016}), ISSN \bibinfo{issn}{0034-4885},
  \urlprefix\url{http://stacks.iop.org/0034-4885/79/i=6/a=064502?key=crossref.ca29f1cdb9c8225a2d270526bf9a3f25}.

\bibitem[{\citenamefont{Schmidt et~al.}(2010)\citenamefont{Schmidt, Hamidian,
  Wahl, Meier, Balatsky, Garrett, Williams, Luke, and Davis}}]{davis2009}
\bibinfo{author}{\bibfnamefont{A.~R.} \bibnamefont{Schmidt}},
  \bibinfo{author}{\bibfnamefont{M.~H.} \bibnamefont{Hamidian}},
  \bibinfo{author}{\bibfnamefont{P.}~\bibnamefont{Wahl}},
  \bibinfo{author}{\bibfnamefont{F.}~\bibnamefont{Meier}},
  \bibinfo{author}{\bibfnamefont{A.~V.} \bibnamefont{Balatsky}},
  \bibinfo{author}{\bibfnamefont{J.~D.} \bibnamefont{Garrett}},
  \bibinfo{author}{\bibfnamefont{T.~J.} \bibnamefont{Williams}},
  \bibinfo{author}{\bibfnamefont{G.~M.} \bibnamefont{Luke}}, \bibnamefont{and}
  \bibinfo{author}{\bibfnamefont{J.~C.} \bibnamefont{Davis}},
  \bibinfo{journal}{Nature} \textbf{\bibinfo{volume}{465}},
  \bibinfo{pages}{570} (\bibinfo{year}{2010}),
  \urlprefix\url{http://dx.doi.org/10.1038/nature09073}.

\bibitem[{\citenamefont{Aynajian et~al.}(2010)\citenamefont{Aynajian,
  da~Silva~Neto, Parker, Huang, Pasupathy, Mydosh, and Yazdani}}]{yazdani2010}
\bibinfo{author}{\bibfnamefont{P.}~\bibnamefont{Aynajian}},
  \bibinfo{author}{\bibfnamefont{E.~H.} \bibnamefont{da~Silva~Neto}},
  \bibinfo{author}{\bibfnamefont{C.~V.} \bibnamefont{Parker}},
  \bibinfo{author}{\bibfnamefont{Y.}~\bibnamefont{Huang}},
  \bibinfo{author}{\bibfnamefont{A.}~\bibnamefont{Pasupathy}},
  \bibinfo{author}{\bibfnamefont{J.}~\bibnamefont{Mydosh}}, \bibnamefont{and}
  \bibinfo{author}{\bibfnamefont{A.}~\bibnamefont{Yazdani}},
  \bibinfo{journal}{Proceedings of the National Academy of Sciences}
  \textbf{\bibinfo{volume}{107}}, \bibinfo{pages}{10383}
  (\bibinfo{year}{2010}).

\bibitem[{\citenamefont{Wray et~al.}(2015)\citenamefont{Wray, Denlinger, Huang,
  He, Butch, Maple, Hussain, and Chuang}}]{Wray2015}
\bibinfo{author}{\bibfnamefont{L.~A.} \bibnamefont{Wray}},
  \bibinfo{author}{\bibfnamefont{J.}~\bibnamefont{Denlinger}},
  \bibinfo{author}{\bibfnamefont{S.-W.} \bibnamefont{Huang}},
  \bibinfo{author}{\bibfnamefont{H.}~\bibnamefont{He}},
  \bibinfo{author}{\bibfnamefont{N.~P.} \bibnamefont{Butch}},
  \bibinfo{author}{\bibfnamefont{M.~B.} \bibnamefont{Maple}},
  \bibinfo{author}{\bibfnamefont{Z.}~\bibnamefont{Hussain}}, \bibnamefont{and}
  \bibinfo{author}{\bibfnamefont{Y.-D.} \bibnamefont{Chuang}},
  \bibinfo{journal}{Physical Review Letters} \textbf{\bibinfo{volume}{114}},
  \bibinfo{pages}{236401} (\bibinfo{year}{2015}), ISSN
  \bibinfo{issn}{0031-9007},
  \urlprefix\url{https://link.aps.org/doi/10.1103/PhysRevLett.114.236401}.

\bibitem[{\citenamefont{Kummer et~al.}(2015)\citenamefont{Kummer, Patil,
  Chikina, G{\"{u}}ttler, H{\"{o}}ppner, Generalov, Danzenb{\"{a}}cher, Seiro,
  Hannaske, Krellner et~al.}}]{Kummer2015}
\bibinfo{author}{\bibfnamefont{K.}~\bibnamefont{Kummer}},
  \bibinfo{author}{\bibfnamefont{S.}~\bibnamefont{Patil}},
  \bibinfo{author}{\bibfnamefont{A.}~\bibnamefont{Chikina}},
  \bibinfo{author}{\bibfnamefont{M.}~\bibnamefont{G{\"{u}}ttler}},
  \bibinfo{author}{\bibfnamefont{M.}~\bibnamefont{H{\"{o}}ppner}},
  \bibinfo{author}{\bibfnamefont{A.}~\bibnamefont{Generalov}},
  \bibinfo{author}{\bibfnamefont{S.}~\bibnamefont{Danzenb{\"{a}}cher}},
  \bibinfo{author}{\bibfnamefont{S.}~\bibnamefont{Seiro}},
  \bibinfo{author}{\bibfnamefont{A.}~\bibnamefont{Hannaske}},
  \bibinfo{author}{\bibfnamefont{C.}~\bibnamefont{Krellner}},
  \bibnamefont{et~al.}, \bibinfo{journal}{Physical Review X}
  \textbf{\bibinfo{volume}{5}}, \bibinfo{pages}{011028} (\bibinfo{year}{2015}),
  ISSN \bibinfo{issn}{2160-3308},
  \urlprefix\url{https://link.aps.org/doi/10.1103/PhysRevX.5.011028}.

\bibitem[{\citenamefont{Fujimori}(2016)}]{Fujimori2016}
\bibinfo{author}{\bibfnamefont{S.-i.} \bibnamefont{Fujimori}},
  \bibinfo{journal}{Journal of Physics: Condensed Matter}
  \textbf{\bibinfo{volume}{28}}, \bibinfo{pages}{153002}
  (\bibinfo{year}{2016}), ISSN \bibinfo{issn}{0953-8984},
  \urlprefix\url{http://stacks.iop.org/0953-8984/28/i=15/a=153002?key=crossref.a4b18b6b6163b9094de0f21343a85c9f}.

\bibitem[{\citenamefont{Patil et~al.}(2016)\citenamefont{Patil, Generalov,
  G{\"{u}}ttler, Kushwaha, Chikina, Kummer, R{\"{o}}del, Santander-Syro,
  Caroca-Canales, Geibel et~al.}}]{Patil2016}
\bibinfo{author}{\bibfnamefont{S.}~\bibnamefont{Patil}},
  \bibinfo{author}{\bibfnamefont{A.}~\bibnamefont{Generalov}},
  \bibinfo{author}{\bibfnamefont{M.}~\bibnamefont{G{\"{u}}ttler}},
  \bibinfo{author}{\bibfnamefont{P.}~\bibnamefont{Kushwaha}},
  \bibinfo{author}{\bibfnamefont{A.}~\bibnamefont{Chikina}},
  \bibinfo{author}{\bibfnamefont{K.}~\bibnamefont{Kummer}},
  \bibinfo{author}{\bibfnamefont{T.~C.} \bibnamefont{R{\"{o}}del}},
  \bibinfo{author}{\bibfnamefont{A.~F.} \bibnamefont{Santander-Syro}},
  \bibinfo{author}{\bibfnamefont{N.}~\bibnamefont{Caroca-Canales}},
  \bibinfo{author}{\bibfnamefont{C.}~\bibnamefont{Geibel}},
  \bibnamefont{et~al.}, \bibinfo{journal}{Nature Communications}
  \textbf{\bibinfo{volume}{7}}, \bibinfo{pages}{11029} (\bibinfo{year}{2016}),
  ISSN \bibinfo{issn}{2041-1723},
  \urlprefix\url{http://www.nature.com/doifinder/10.1038/ncomms11029}.

\bibitem[{\citenamefont{Cardona}(1983)}]{Cardona1983}
\bibinfo{author}{\bibfnamefont{M.}~\bibnamefont{Cardona}},
  \emph{\bibinfo{title}{{Light Scattering in Solids I : Introductory
  Concepts}}} (\bibinfo{publisher}{Springer Berlin Heidelberg},
  \bibinfo{year}{1983}), ISBN \bibinfo{isbn}{9783540707554}.

\bibitem[{\citenamefont{Severing et~al.}(1990)\citenamefont{Severing, Gratz,
  Rainford, and Yoshimura}}]{Severing1990}
\bibinfo{author}{\bibfnamefont{A.}~\bibnamefont{Severing}},
  \bibinfo{author}{\bibfnamefont{E.}~\bibnamefont{Gratz}},
  \bibinfo{author}{\bibfnamefont{B.}~\bibnamefont{Rainford}}, \bibnamefont{and}
  \bibinfo{author}{\bibfnamefont{K.}~\bibnamefont{Yoshimura}},
  \bibinfo{journal}{Physica B: Condensed Matter}
  \textbf{\bibinfo{volume}{163}}, \bibinfo{pages}{409 } (\bibinfo{year}{1990}),
  ISSN \bibinfo{issn}{0921-4526}.

\bibitem[{\citenamefont{Kotani}(2011)}]{kotani11}
\bibinfo{author}{\bibfnamefont{A.}~\bibnamefont{Kotani}},
  \bibinfo{journal}{Phys. Rev. B} \textbf{\bibinfo{volume}{83}},
  \bibinfo{pages}{165126} (\bibinfo{year}{2011}),
  \urlprefix\url{http://link.aps.org/doi/10.1103/PhysRevB.83.165126}.

\bibitem[{\citenamefont{Varma and Yafet}(1976)}]{Varma1976}
\bibinfo{author}{\bibfnamefont{C.~M.} \bibnamefont{Varma}} \bibnamefont{and}
  \bibinfo{author}{\bibfnamefont{Y.}~\bibnamefont{Yafet}},
  \bibinfo{journal}{Phys. Rev. B} \textbf{\bibinfo{volume}{13}},
  \bibinfo{pages}{2950} (\bibinfo{year}{1976}).

\bibitem[{\citenamefont{Batlogg et~al.}(1976)\citenamefont{Batlogg, Kaldis,
  Schlegel, and Wachter}}]{batlogg76}
\bibinfo{author}{\bibfnamefont{B.}~\bibnamefont{Batlogg}},
  \bibinfo{author}{\bibfnamefont{E.}~\bibnamefont{Kaldis}},
  \bibinfo{author}{\bibfnamefont{A.}~\bibnamefont{Schlegel}}, \bibnamefont{and}
  \bibinfo{author}{\bibfnamefont{P.}~\bibnamefont{Wachter}},
  \bibinfo{journal}{Phys. Rev. B} \textbf{\bibinfo{volume}{14}},
  \bibinfo{pages}{5503} (\bibinfo{year}{1976}),
  \urlprefix\url{http://link.aps.org/doi/10.1103/PhysRevB.14.5503}.

\end{thebibliography}
\end{document}